\documentclass[aps,showpacs,preprint,nofootinbib]{revtex4}
\begin{document}

\title{On Hamiltonian formulation of the Einstein-Hilbert action in two 
dimensions}
\author{N. Kiriushcheva and S.V. Kuzmin}
%\email{nkiriush@uwo.ca}
%\affiliation{Department of Applied Mathematics,}
%\author{S.V. Kuzmin}
\email{nkiriush@uwo.ca, skuzmin@uwo.ca}
\affiliation{Department of Applied Mathematics, 
University of Western Ontario, London, N6A~5B7 Canada}
 
\date{\today}

%\maketitle

\begin{abstract}
It is shown that the well-known triviality of the Einstein field 
equations in two dimensions is not a sufficient condition for the 
Einstein-Hilbert action to be a total divergence, if the general 
covariance is to be preserved, that is, a coordinate system is not fixed. 
Consequently, a Hamiltonian formulation is possible without any modification 
of the two dimensional Einstein-Hilbert action. We find the resulting 
constraints and the corresponding gauge transfromations of the metric tensor. 

%{\it PACS:} 11.10.Ef

{\it Keywords:} Hamiltonian formulation, Einstein-Hilbert action, two 
dimensions.
\pacs{11.10.Ef}
\end{abstract}

\maketitle

In this letter we reconsider the canonical formulation of the two dimensional 
Einstein-Hilbert (EH) action. It is generally believed that 
the Einstein Lagrangian is a total derivative in two dimensions ($2D$) 
\cite{Polchinski, Mann} and that its canonical formulation  
cannot be constructed. (In \cite {Martinec} it is stated that 
``the canonical formalism breaks down'' and in \cite {Strobl} we find that in
two spacetime dimensions EH action is meaningless and its 
modification is needed. It is true that if the Lagrangian 
$\sqrt{-g}R$ is a total divergence and it is dropped, 
then the remaining cosmological term in the Lagrangian $-\Lambda\sqrt{-g}$
contains 
no derivatives and the canonical momenta cannot be defined \cite{Banks}.)
However, this statement deserves a closer examination as it seems to 
contradict some general principles. The Einstein-Hilbert action is  
valid in any dimension and, of course, can have specific behavior in a 
particular dimension, but the absence of a canonical formulation would be 
similar to the claim that an action has no equations of motion in some 
dimension which is incorrect (equations of motion can be trivial,  
but they do exist). In such a situation it  
is natural to expect that the Hamiltonian formulation of a theory  
can be trivial, leading to zero degrees of freedom, but it does exist.
Moreover, if an action is a total derivative (in covariant form) in a 
particular dimension it has to be a total derivative in all dimensions 
which is not the case for the Einstein-Hilbert action.

The source of such belief is originated from the well-known fact 
(\cite {LL}, \cite {Carmeli}) that the Einstein equations 
\begin{equation}
\label{1}R_{\mu \nu}-\frac{1}{2}g_{\mu \nu}R=0
\end{equation}
are trivial in $2D$. This is easy to 
demonstrate using the fact that
only one component of Riemann tensor is independent in $2D$ \cite {LL}. 
It can also be shown 
by straightforward variation of the EH action (we use the 
signature $\left(+,-,-,...\right)$):
\begin{equation}
\label{2}S_{EH}=\int d^Dx \sqrt{\left(-1\right)^{D-1}g} g^{\mu \nu } 
R_{\mu \nu },
\end{equation}
with respect to metric tensor $g_{\mu \nu}$. (Here 
$g=\det{\left(g_{\mu \nu }\right)}$, 
$R_{\mu \nu }$ is Ricci tensor
$
R_{\mu \nu }=\Gamma_{\mu \nu ,\lambda}^\lambda -\Gamma _{\mu \lambda ,\nu }
^\lambda +\Gamma _{\sigma \lambda
}^\lambda \Gamma _{\mu \nu }^\sigma -\Gamma _{\sigma \mu }^\lambda
\Gamma _{\nu \lambda }^\sigma $
and $\Gamma _{\mu \nu }^\sigma$ is Christoffel symbol
$\Gamma _{\mu \nu }^\sigma = \frac{1}{2} g^{\lambda \sigma } \left( g_{\mu 
\sigma , \nu } + 
g_{\nu \sigma , \mu } - g_{\mu \nu , \sigma } \right)$.)

However, the 
conjecture that ``correspondingly the EH action is a surface term'' 
\cite {Jackiw}, based on the fact that the equations of motion are trivial, 
is not
obvious and has to be checked. The triviality of equations of motion is a  
necessary condition for the Lagrangian to be a total divergence but not a  
sufficient one. We shall now show that the EH action in two dimensions 
provides a counter example.

The extraction of the terms in the EH action that can be cast into a 
total divergence is well-known in any dimension \cite {LL}. 
This separation has been  
used in attempts to devise Hamiltonian formulation of General 
Relativity (GR) \cite {Pirani}, \cite {Dirac}. We can write the action of 
(\ref{2}) as
\begin{equation}
\label{3}S_{EH}= \int d^Dx L_{\Gamma \Gamma}-\int d^Dx \left(V^{\alpha} 
 \right)_{,\alpha}
\end{equation}
where the ``Gamma-Gamma'' part is
\begin{equation}
\label{4}L_{\Gamma \Gamma } = \sqrt{\left(-1\right)^{D-1}g} g^{\mu \nu }
\left(\Gamma _{\sigma \mu }^\lambda \Gamma _{\nu \lambda }^\sigma -
\Gamma _{\sigma \lambda}^\lambda \Gamma _{\mu \nu }^\sigma\right)
\end{equation}
and $V^\alpha$ is
\begin{equation}
\label{5}V^\alpha = \sqrt{\left(-1\right)^{D-1}g}
\left(g^{\alpha \mu} \Gamma_{\mu \nu}^\nu - g^{\mu \nu} 
\Gamma_{\mu \nu}^\alpha \right)= \sqrt{\left(-1\right)^{D-1}g}
\left[g^{\alpha \mu}g^{\nu \beta}\left(
g_{\nu \beta, \mu} - g_{\nu \mu, \beta}\right)\right].
\end{equation}

After extracting the total divergence, the Jackiw conjecture 
\cite {Jackiw} has to lead to zero Gamma-Gamma part in the $2D$ action. 
We can, however, show that this is incorrect. We express the Gamma-Gamma term 
in the form (see Eq. (2) of \cite{Pirani} or Eq. (8) 
of \cite {Dirac})
\begin{equation}
\label{6}L_{\Gamma \Gamma}=\frac{1}{4} \sqrt{\left(-1\right)^{D-1}g} 
G^{\left(\mu \nu, \rho\right)\left(\alpha \beta, \sigma\right)}
g_{\mu \nu, \rho} g_{\alpha \beta, 
\sigma}
\end{equation}
where
\begin{equation}
\label{7}G^{\left(\mu \nu, \rho\right)\left(\alpha \beta, \sigma\right)}=
\left(g^{\mu \nu}g^{\alpha \beta} - g^{\mu \alpha} g^{\nu \beta}
\right)g^{\rho \sigma} + 2 \left(g^{\mu \alpha}g^{\beta \rho} - g^{\mu \rho}
g^{\alpha \beta}\right)g^{\nu \sigma}.
\end{equation}
(Eq. (\ref {7}) in \cite {Dirac} has the opposite sign, but we keep 
convention of \cite {LL}.)

In $2D$, there are no contributions with two temporal derivatives. This  
immediately follows from Eq.(9) of \cite{Dirac}:
\begin{equation}
\label{8}g_{ik,0}g_{ml,0} G^{\left(ik,0\right)\left(ml,0\right)}= 
g_{ik,0}g_{ml,0}g^{00}\left(e^{im}e^{kl}-e^{ik}e^{ml} \right)
\end{equation}
where $e^{ik}=g^{ik}-\frac{g^{0i}g^{0k}}{g^{00}}$.
(Latin indices indicate spatial components.)

Similarly, the contributions to (\ref{6}) that have two spatial 
derivatives vanish in $2D$,
\begin{equation}
\label{9}\left[g_{\mu \nu, k}g_{\alpha \beta, m} 
G^{\left(\mu \nu, k\right)\left(\alpha \beta, m\right)}\right]_{D=2}=0.
\end{equation}

However, some cross-terms with both spatial and temporal derivatives do not 
cancel and can be presented in the following, ``semi-covariant'', form 
\begin{equation}
\label{10}L_{\Gamma \Gamma}= \frac{1}{2} \sqrt{-g} e^{11}g^{00}
g^{\alpha \beta}\left(
g_{0 \alpha, 1}g_{\beta 1, 0} - g_{0 \alpha, 0}g_{\beta 1, 1}\right).
\end{equation}

It is not difficult to perform variation of (\ref{10}) with respect to 
the metric$g_{\alpha \beta}$. This results in 
trivial equations of motion , consistent with what is a well-known 
result. However (\ref {10}) is not equal to zero identically 
and cannot be put in the form of a total derivative, contrary to the generally 
held belief, as it was written in \cite {Jackiw} 
and in many other articles including 
our own \cite {KKM} where we thoughtlessly repeated this.
Unfortunately, we also refer to \cite{LL} as a proof of it, which is
entirely our mistake (we would like to note that 
L.D. Landau and E.M. Lifshits do provided a proof of the triviality of 
Einstein equations in $2D$ but they had never made the conjecture that 
triviality of the equations of motion is equivalent to the Lagrangian 
being a total divergence). Of course, (\ref {10}) 
can be made equal to zero by choice of a particular coordinate
system or a subset of coordinate systems, but this contradicts 
the Dirac procedure \cite {Diracbook} for passing to a Hamiltonian 
formulation for gauge theories. The main tenet of this procedure is 
to avoid any reference to a particular coordinate system, because 
when using a Hamiltonian 
general formulation, gauge transfromations can be restored from 
a knowledge of the first class constraints which should be independent of any 
choice of coordinate system. 
 
One simple example in which (\ref{10}) vanishes is the synchronous 
coordinate system (or ``synchronous gauge'' \cite {Banks}) in which
\begin{equation}
\label{10a}
g_{0k}=0, g_{00}=1
\end{equation}
and time lines are normal to the hypersurfaces $x_0 =const$ (see Sec. 97 
\cite {LL}). Under condition (\ref {10a}), equation (\ref {10}) is zero and
$V^\alpha_{,\alpha}$ in 
(\ref {3}) is equal to $2\left(\sqrt{-g_{11}}\right)_{,0,0}$. 

Similarly, the coordinate transfromations 
\begin{equation}
\label{10b}g_{\mu \nu}=e^{\phi}\eta_{\mu \nu}
\end{equation}
reduces (\ref {10}) to zero and  $V^\alpha_{,\alpha}$ becomes  
$\partial_\mu \partial^\mu \phi$ (i.e., see p. 304 of \cite{string}).  

In \cite{Strobl} the proof that $\sqrt{-g} R$ is a total divergence in 
Cartan variables was given using an {\it orthonormal basis}. Any choice of  
variables which leads to $g_{01}=0$ (and, consequently, $g^{01}=0$)
makes (\ref {10}) equal to zero. Not surprisingly,    
in such cases, the canonical formalism ``breaks down'' \cite{Martinec} and
the action is meaningless \cite{Strobl}.
If some reformulation of the EH action reduces the action to 
a total divergence in $2D$, it means that general covariance
in such formulation has been destroyed.  

If we keep the action covariant (without specifying a coordinate system)
then a canonical formulation of (\ref {10}) can be performed (contrary to 
the conjecture of \cite {Martinec}). The Lagrangian (\ref {10}) is very 
simple, and passing to a Hamiltonian form is straightforward since 
all terms are linear in velocities and introduction of momenta 
$\pi^{\alpha \beta}$ conjugate to metric tensor $g_{\alpha \beta}$
leads immediately to three primary constraints \cite{Diracbook}
$$
\phi^{00}=\pi^{00} + \frac{1}{2} \sqrt{-g} g^{00}g^{11}
\left(g^{00}g_{01,1} + g^{01}g_{11,1}\right),
$$
\begin{equation}
\label{11}\phi^{01}=\pi^{01} - \frac{1}{4} \sqrt{-g} g^{00}g^{11}
\left(g^{00}g_{00,1} - g^{11}g_{11,1}\right),
\end{equation}
$$
\phi^{11}=\pi^{11} - \frac{1}{2} \sqrt{-g} g^{00}g^{11}
\left(g^{01}g_{00,1} + g^{11}g_{01,1}\right).
$$
 
The Hamiltonian is then just a linear combination of constraints (\ref {11})
\begin{equation}
\label{12}H=\lambda_{\alpha \beta} \phi^{\alpha \beta}.
\end{equation}

Using the fundamental Poisson brackets (PB)
\begin{equation}
\label{14}\left\{g_{\alpha \beta},\pi^{\mu \nu} \right\}=
\frac{1}{2}\left(\delta_\alpha^\mu \delta_\beta^\nu +
\delta_\alpha^\nu \delta_\beta^\mu\right),
\end{equation} 
it is easy to demonstrate that each of the constraints has a vanishing 
PB with itself as expected because of the 
antisymmetric properties of the PB. (This is unlike the
hypersurface deformation algebra of the ADM constraints \cite{Thiemann}.) 
This is actually 
obvious because in all constraints appearing in (\ref{11}) there are 
no spatial derivatives of the  
components of $g_{\alpha \beta}$ which correspond to components of the  
conjujate momenta $\pi^{\alpha \beta}$. Moreover, the PB's among different
constraints are also zero and the resulting algebra is quite trivial:
\begin{equation}
\label{15}\left\{\phi^{\alpha \beta},\phi^{\mu \nu} \right\}=
\frac{\delta \phi^{\alpha \beta}}{\delta g_{\rho \sigma}}
\frac{\delta \phi^{\mu \nu}}{\delta \pi^{\rho \sigma}}-
\frac{\delta \phi^{\mu \nu}}{\delta g_{\rho \sigma}}
\frac{\delta \phi^{\alpha \beta}}{\delta \pi^{\rho \sigma}}=0.
\end{equation}

From (\ref{12}, \ref{15}) it is obvious that there are no 
secondary constraints. The Dirac procedure is closed; standard counting of 
degrees of freedom leads to zero as there are three first class (FC) 
constraints corresponding to the three independent components of the metric 
tensor.

From these three primary FC constraints we can construct the gauge generator 
$G\left(\epsilon\right)$ using the Castellani procedure \cite{Castellani}. 
In this case we simply get
\begin{equation}
\label{16}G\left(\epsilon\right) = \int dx \epsilon_{\alpha \beta} 
\phi^{\alpha \beta}
\end{equation}
where the $\epsilon_{\alpha \beta}$ are gauge parameters.

The gauge transformations of $g_{\alpha \beta}$ can be found from 
$\delta g_{\alpha \beta} = \left\{g_{\alpha \beta},G\left(\epsilon\right)
\right\}$ giving,
\begin{equation}
\label{17}\delta g_{\alpha \beta} = \epsilon_{\alpha \beta}.
\end{equation}
This is expected as the equations of motion are trivial equations which 
are valid for any $g_{\alpha \beta}$.

It is easy to check that these gauge transformations leave $S_{EH}$ 
invariant up to a surface term, because the variation of 
$L_{\Gamma \Gamma}$ is:

\begin{equation}
\label{18}\delta L_{\Gamma \Gamma} = \frac{1}{2}\partial_\gamma\left[\sqrt{-g}
\left(g^{\rho \sigma}g^{\beta \gamma} - g^{\rho \gamma}g^{\beta \sigma}\right)
g^{\alpha \mu}g_{\mu \rho, \sigma} \epsilon_{\alpha \beta}\right].
\end{equation}

We have demonstrated that the statements that $2D$ EH action is meaningless 
and that canonical procedure for it breaks down are not in fact correct and 
are just a consequence of not following canonical procedure by 
having made the choice of some ``priviledged'' coordinate system or a family 
of such systems including ``slicing of spacetime''. (According to Hawking 
\cite{Hawking} this also contradicts to the spirit of GR.) Does any meaning
exist in ``canonical'' formulations which results depend on a particular 
coordinate system and in some dimensions cannot be even formulated? 
Of course, it is possible to use a 
specific coordinate system to study particular classical solutions of the
Einstein field equations, but it is inappropriate to make such a choice when 
quantizing GR, as quantum
fluctuations are not restricted to a particular choice of coordinates.
If the canonical procedure is performed 
in a fixed coordinate system, we cannot guarantee that it will reproduce 
the same invariance as that present in the original action. In $2D$ the use 
of a particular system does not even allow one to perform a 
canonical procedure. If we want to 
ensure that the Quantum Gravity is consistent with GR we have to retain the 
principle of general covariance when applying canonical procedure and not  
destroying general covariance from the beginning.  
The simple example of the EH action in $2D$ shows the importance 
of keeping  general covariance when using the canonical procedure.

Invariance of the Lagrangian up to a surface term is different from the exact 
invariance occuring in ordinary gauge theories and one needs to impose an 
extra condition on behavior of gauge parameters at infinity as, for example, 
in the case of invariance of the Gamma-Gamma part of the EH action under 
linearized coordinate transformations 
(see p. 272 of \cite {LL}). In particular, we are not aware of any canonical 
formulation of the $2D$ EH action that restores these transformations. 
The use of the ADM 
formulation \cite{ADMlast} for the $2D$ EH action  leads to the unphysical 
result of there being  negative degrees of freedom \cite{Martinec}; i.e. it 
is an overconstrained system. 
Contrary, our (\ref{17}) is the result of canonical procedure for the 
non-divergence part of the $2D$ EH action.

However, we think that invariance up to a surface term as given by (\ref {18}) 
could be the result of another deviation from general covariance, 
not peculiarity of EH action. The reason for this is the elimination of a 
total divergence in our consideration and actually
taking only part of EH action, $S_{\Gamma \Gamma}$-part, which is not 
covariant (see \cite {Carmeli}, \cite{PiraniShild}). To  
obtain Hamiltonian formulation which can lead to 
gauge transfromations of the same transforming power as 
general coordinate transformations of the EH action (not only 
$S_{\Gamma \Gamma}$) we have to keep the effect of all terms in the course of 
canonical procedure (for discussion of this point see \cite {KKM3}). 
In higher dimensions one possibility is to use the equivalent 
first order affine-metric formulation of Einstein \cite{Einstein1925}, which, 
however, cannot be used in $2D$, as affine 
connections in this case cannot be found in terms of the metric tensor 
\cite {Mann}.
This is also true for its particular combination used in \cite {KKM3} and, 
actually, for any linear combination of them. 

So, a first order formulation different from the affine-metric one has to be 
used in a canonical approach to the $2D$ EH action. There exists a variety
of first order formulations (non-symplectic)
that can be constructed keeping equivalence with the second order form.
The question is whether one of them can give consistent Hamiltonian 
formulation and what gauge transformations it will produce.

We can also try to capture the 
effect of all terms in the EH action performing the canonical procedure by 
considering the EH action as a theory with higher derivatives and using 
Ostrogradsky method \cite{Ostrogradsky} with modifications appropriate 
for singular systems 
\cite{GT, GTbook}. (The first attempt to apply this to the full EH action 
is due to Dutt and Dresden \cite{DD}.)  

We would like to conclude by referring to a conjecture of Dirac 
\cite{Diracbook}:
{``\it I} (Dirac) {\it feel that there will always be something missing 
from them} 
(non-Hamiltonian methods) {\it which we can only get by working from a 
Hamiltonian''}.
This article provides an example that is an 
illustration of this conjecture; despite triviality of 
equations of motion, the Hamiltonian formulation allows one to find a gauge 
transformation and the corresponding invariance of the action.

%\section{Acknowledgments}
{\bf ACKNOWLEDGMENTS}

The authors are greateful to D.G.C. McKeon for discussions and reading the 
manuscript. We thank E.V. Gorbar for reading of the final version of 
manuscript, S.B. Gryb for discussions, D.V. Prokopiev and E.Z. Prokopieva 
for useful suggestions and R. Corless for inspiration and support.

\end{document}